\documentclass[12pt]{iopart}

\usepackage[utf8]{inputenc}
\usepackage{hyperref}
\hypersetup{  
colorlinks=true,
citecolor=blue} 
\usepackage{iopams} 
\usepackage{bm}
\usepackage{graphicx}

\newcommand{\f}[2]{{\frac{#1}{#2}}}
\newcommand{\text}{\mathrm}

\newcommand{\I}{1}
\newcommand{\II}{2}

\newcommand{\X}{{\scriptscriptstyle{X}}}
\newcommand{\Y}{{\scriptscriptstyle{Y}}}
\newcommand{\exy}{{(x,y)}}
\newcommand{\eyx}{{(y,x)}}
\newcommand{\ome}{{x}}
\renewcommand{\oe}{{y}}
\newcommand{\geoact}{g}

\newcommand{\act}{\bar A}
\newcommand{\T}{\mathrm{T}}
\newcommand{\vect}[2][]{\bm{#2}_{\mathrm{#1}}}
\newcommand{\Jmoyi}[1]{J_{#1}}
\newcommand{\Jstochi}[1]{j_{#1}}

\newcommand{\Aphi}{A_\phi}
\newcommand{\be}{\begin{equation}}
\newcommand{\ee}{\end{equation}}
\newcommand{\ba}{\begin{eqnarray}}
\newcommand{\ea}{\end{eqnarray}}
\newcommand{\eqref}[1]{(\ref{#1})}
\renewcommand{\geq}{\geqslant}
\renewcommand{\leq}{\leqslant}

\begin{document}

\title{An ordered set of power-efficiency trade-offs}

\author{Hadrien VROYLANDT$^{1,2}$, David LACOSTE$^3$, Gatien VERLEY$^1$}
\address{$^1$ Laboratoire de Physique Théorique (UMR8627), CNRS, Univ. Paris-Sud, Université Paris-Saclay, 91405 Orsay, France}
\address{$^2$Department of Chemistry, Northwestern University, Evanston, IL 60208}
\address{$^3$ Gulliver Laboratory, UMR CNRS 7083, PSL Research University, 
ESPCI, 10 rue Vauquelin, F-75231 Paris,
France}

\submitto{\JSTAT}

\begin{abstract}
In this paper, we derive a number of inequalities which express power-efficiency trade-offs that hold generally
for thermodynamic machines operating in non-equilibrium stationary states. 
One of these inequalities concerns the output power, which is bounded by a quadratic function 
of the thermodynamic efficiency multiplied by a factor. 
Different factors can be obtained according to 
the level of knowledge one has about the underlying dynamics of the machine, they can depend 
for instance on the covariance of the input flux, the dynamical activity, or the non-equilibrium conductance.
\end{abstract}

\section*{Introduction}
\label{sec:Intro}

In recent years, considerable efforts have been devoted to 
engineer new thermoelectric materials with the best possible efficiency \cite{Benenti2017_vol694} and 
to build small artificial stochastic engines mimicking molecular motors \cite{Bustamante2005_vol58, Porto2000_vol84,Blickle2012_vol8}. Clearly, in order to build the best possible machines, it is essential to develop a general understanding of the relationship between power, precision and dissipation \cite{Pietzonka2018_vol120}. What are the fundamental limits and design trade-offs 
involved in optimizing these three quantities ?  

This question is related to a major recent development in Stochastic Thermodynamics called 
the thermodynamic uncertainty relation, which is important because it goes beyond the usual 
formulation of the second law of thermodynamics \cite{Dechant2018}. 
This result establishes that the precision on a thermodynamic current in non-equilibrium stationary states
comes with a minimal energetic cost \cite{Barato2015,Gingrich2016_vol116}, 
where precision is quantified by the variance of the current 
and the energetic cost is measured by the dissipation.
Applications of this thermodynamic uncertainty relation include among others, 
an inference method to obtain the topology or the dissipation present in chemical networks 
\cite{Proesmans2018,Gingrich2017a,Barato2015d}, a characterization of brownian clocks \cite{Barato2016}, 
bounds on the efficiency of molecular motors \cite{Pietzonka2016b}, 
design principles on non-equilibrium self-assembly \cite{Nguyen2016} and much more.

For stochastic dynamics in contact with heat baths, a related result derived by Shiraishi et 
al. \cite{Shiraishi2016} states that the square of the heat current between the system and heat bath 
is bounded by a system-dependent positive constant times the rate of entropy production. 
The Shiraishi et al. result and the thermodynamic uncertainty relation both lead to similar 
power-efficiency trade-offs as far as the dependence on efficiency is concerned and the main difference 
between the two results lies in a system-dependent constant in factor of the function of the efficiency. 
Regardless of the precise value of this system-dependent positive constant, 
both results imply that the maximal efficiency of machines can only be realized
 at vanishing power output. The similarity between these two formulations of the power-efficiency 
trade-offs suggests that a general framework could exist, which presumably would include both formulations 
in a unifying way. 

The search for such an unifying framework is motivating the present paper. In fact, a number of recent works are going 
in this direction: on one hand, the result of Shiraishi et al. has been generalized to arbitrary currents 
besides the heat current, for non-thermal heat baths, and for dynamics with broken time-reversal symmetry 
but keeping the assumption of Langevin dynamics \cite{Dechant2018}. 
These authors obtained a general inequality based on the Cauchy-Schwartz inequality, according to which, 
the rate of entropy production is bounded from below by the square 
of any irreversible current. On the other hand, some of the limitations of the thermodynamic uncertainty relation 
have now been overcome, such as the assumption of steady states. Indeed, in Ref. \cite{Barato2018}
time-periodic machines have been studied in this context. 
These new results also follow from bounds on large deviation functions of a single
 current as in the original uncertainty relation, except that they no longer involve the
 entropy production, which is replaced by a different quantity. This quantity can be interpreted 
as the entropy production of the stationary dynamics that has the same mean current. 
Finally, another limitation of the uncertainty relation, the requirement
 of not breaking time-reversal dynamics, has been addressed in Ref.~\cite{Macieszczak2018}.

In this paper, we follow a somewhat different route as compared to these works, while still aiming at unifying 
power-efficiency trade-offs. Our approach is based on a concept we introduced in an earlier work, namely 
that of non-equilibrium conductance matrix \cite{Vroylandt2018}.
This conductance matrix, relates physical currents to thermodynamic forces, just like the Onsager matrix, 
but generalizes it by being not limited to the near equilibrium regime. This new framework holds for
 systems operating in general non-equilibrium stationary states, i.e. arbitrarily far from equilibrium. 
By construction, this conductance matrix is a real, symmetric and semi-definite positive matrix, just like the
 Onsager matrix. One important difference with the Onsager matrix however,
 is that the coefficients of this matrix are not constants, 
but are functions of thermodynamic forces. Only near equilibrium, this dependence can be neglected 
in which case the non-equilibrium conductance matrix becomes identical with the Onsager matrix. 
This similarity with the Onsager matrix, allowed us to prove that the maximum thermodynamic efficiency achievable 
by a thermodynamic machine only depends of the so-called degree of coupling of the thermodynamic machine 
\cite{Vroylandt2018}, 
thus generalizing an old result which was known for machines operating near equilibrium \cite{Caplan1966_vol10}. 
We also noted that the macroscopic current-force relation does not lead to a unique conductance 
matrix, while a unique matrix can be built if the microscopic dynamics is known. 
To obtain an explicit matrix in this way, we considered a dynamics of Markov jump processes,
and we obtained the non-equilibrium conductance matrix by extending 
a previously introduced large deviation formalism of stochastic currents \cite{Polettini2016_vol94a}.

In this paper, we derive a number of bounds using the method introduced in Ref.~\cite{Vroylandt2018} and 
we make contact with the results of Dechant and Sasa\cite{Dechant2018} and of Shiraishi \textit{et al.}\cite{Shiraishi2016}. 
 We find a hierarchy of inequalities in terms of either the conductance matrix, an activity matrix (which is a
 variant of the conductance matrix built from the transition frequencies instead of the local resistances), and 
the covariance matrix of the physical currents. This hierarchy of inequalities represents a generalization of the
 thermodynamic uncertainty relation that naturally leads to power-efficiency trade-offs. Finally, we illustrate
 these trade-offs using two examples of thermodynamic machines.

\section{Power-efficiency trade-offs}
\label{general-discussion}

\subsection{Bounds on the output power}

Let us focus on the simple case of a machine, in which a driving process, which we call the first process, 
drives another process, the second process. 
If we call $\sigma_\I$ (resp. $\sigma_\II$) the partial entropy production rate of the first (resp. second) 
process,
we have $\sigma_\I \geq 0$ and $\sigma_\II \leq 0$. Let us then define the total entropy production as 
$\sigma=\sigma_\I + \sigma_\II$, and the thermodynamic efficiency as $\eta = -\sigma_{\II}/\sigma_{\I}$.  
Using the definition of $\eta$ and the second law of thermodynamics $\sigma \ge 0$, we have $ 1 \geq \eta \geq 0$.

Let us also denote $F_i$ the affinity and $J_i$ the corresponding physical current of 
the process $i=1,2$ of the machine, then 
the partial entropy production $\sigma_i$ is simply $\sigma_i=F_i J_i$.
As explained above, we relate the physical currents to the affinities 
by a generalization of the Onsager matrix, which we call the non-equilibrium conductance matrix $\bm{G}$, 
in such a way that $J_{\X} = \sum_{\Y} G_{\X,\Y} F_{\Y}$ \cite{Vroylandt2018}. 
We then introduce a new parametrization of this matrix in terms of the degree of coupling 
$\xi = G_{12}/ \sqrt{G_{11}G_{22}} \times \mathrm{sign}\, (F_{1} F_{2})$ and the relative intrinsic 
dissipation $\varphi = \sqrt{(G_{22} F_2^2)/ (G_{22}F_{1}^{2})}$.
By expressing the output power $-\sigma_\II$ in terms of these parameters
and optimizing with respect to them, 
we obtain the power-efficiency inequality :
\begin{equation}
  \label{eq:PowerEffS1}
-\sigma_{\II} \leq  G_{1,1} F_1^2 \eta (1-\eta),
\end{equation}
and alternatively using the component $G_{\II,\II}$ of the non-equilibrium conductance matrix 
\begin{equation}   \label{eq:PowerEffS2}
	-\sigma_{\II}  \leq G_{2,2} F_2^2 \f{1-\eta}{\eta}.
\end{equation}

An interesting and important consequence of these inequalities 
is that the output power (proportional to $-\sigma_{\II}$) must vanish when the efficiency approaches
 its maximum  value, {\it i.e.} when $\eta \to 1$, which corresponds for heat engines to the Carnot efficiency, 
unless both coefficients $G_{\I,\I} F_1^2 $ or $G_{2,2} F_2^2$ diverge. This rather unusual limit has been
 considered in Ref.~\cite{Polettini2017_vol118,Apertet2017_vol120}.

An inequality of the type of Eq.~\eqref{eq:PowerEffS1} has been first derived in Ref. \cite{Shiraishi2016} 
for heat engines. In that work, the coefficient $G_{\I,\I} F_1^2$ was replaced by a model dependent coefficient
 $\bar{\Theta}$, for which an expression was provided for a system interacting with Langevin heat baths, 
in terms of the time average of the total kinetic energy of the engine, the temperature (of the baths), 
mass (of the engine) and damping constant (of the engine). A similar inequality has been derived in 
Ref.~\cite{Pietzonka2016b} by P. Pietzonka et al. in the context of molecular motors based on the
 thermodynamic uncertainty relations \cite{Barato2015,Gingrich2016_vol116}. In their case,
 $G_{\I,\I} F_1^2$ is replaced by the variance of the input current.

\subsection{Bounds on the input power}
A similar calculation as that used to derive Eqs. \eqref{eq:PowerEffS1}-\eqref{eq:PowerEffS2} also gives
 bounds on the input power $\sigma_1$ and on the total entropy production $\sigma$. 
Two types of bounds can 
be obtained by making the process one or two special. If one chooses to specialize to the process one, the 
input power $\sigma_1$ takes the following expression : 
\be
\label{sigma1Ineq}
\sigma_1= F_1^2 G_{11} \left( 1 + \xi \varphi \right).
\ee
By optimizing this expression with respect to $\varphi$ at constant $\xi$, one obtains a lower bound which only depends on the degree of coupling :
\be
\label{sigma1}
\sigma_1 \geq F_1^2 G_{11} \left( 1 - \xi^2 \right).
\ee
As also done in the derivation of Eqs. \eqref{eq:PowerEffS1}-\eqref{eq:PowerEffS2}, in this optimization, 
one can treat $G_{11}$ as constant, because there are only two independent parameters
 in the conductance matrix, so they can be chosen to be $\varphi$ and $\xi$.

In order to obtain a different bound now in terms of the efficiency $\eta$ rather than the degree of coupling, 
one uses the expression of $\varphi$ as a function of $\eta$ and $\xi$ \cite{Vroylandt2018} : 
\be
\varphi^\pm = - \frac{\xi \left( \eta + 1 \right)}{2} \pm \frac{1}{2} \sqrt{(\eta + 1)^2 \xi^2 - 4 \eta},  
\ee  
which is then reported into Eq. \eqref{sigma1Ineq}. One obtains two functions of $\xi$, $\sigma_1^\pm (\xi)$,  which are such that 
$\sigma_1^+(\xi) \geq \sigma_1^-(\xi)$. Since $\sigma_1^+(\xi)$ is a monotonously decreasing function of $\xi$, this function reaches its maximum at $\xi=-1$. Reporting this value into the expression of $\sigma_1^+$ leads to the upper bound
\begin{equation}
\label{ex-eq3}
	\sigma_{1} \leq G_{11} F^{2}_{1}(1-\eta).
\end{equation}

If we instead choose to make the second process special, one starts with
\be
\sigma_1= F_2^2 G_{22} \frac{1 + \xi \varphi }{\varphi^2}.
\ee
Now, after reporting the expression of $\varphi^\pm$ into this $\sigma_1$, one 
obtains two solutions which are such that $\sigma_1^+(\xi) \leq \sigma_1^-(\xi)$.
Then, the upper bound is obtained by reporting $\xi=-1$ into $\sigma_1^-$, which leads to
\begin{equation}
\label{ex-eq4}
	\sigma_{1} \leq G_{22} F^{2}_{2}\frac{1-\eta}{\eta^{2}}.
\end{equation}

\subsection{Bounds on the total entropy production}
Similarly, the total entropy production can be expressed in terms of $\varphi$ and $\xi$ by 
choosing either the first or the second process as special. An optimization with respect to 
$\varphi$ at constant $\xi$ leads in the former case to the bound :
\be
\label{eq:boundsEP11}
\sigma \geq F_{1}^2 G_{11} \left(1 - \xi^2 \right),
\ee
and to
\be
\label{eq:boundsEP22}
\sigma \geq F_{2}^2 G_{22} \left(1 - \xi^2 \right),
\ee
in the later case.
It is interesting to note that these lower bounds represent an improvement 
with respect to the second law, except at tight coupling when $\xi=-1$ where the 
inequalities \eqref{eq:boundsEP11}-\eqref{eq:boundsEP22} become 
 the second law $\sigma \geq 0$.
Similarly, for the partial entropy production, $\eqref{sigma1}$
represents an improvement with respect to the second law for the partial entropy production $\sigma_1 \geq 0$
except at tight coupling.
Interestingly, in addition to these lower bounds, this framework also leads to upper bounds on the input power 
such as \eqref{ex-eq3},\eqref{ex-eq4}. In the limit where $\eta \to 1$, these upper bounds 
impose that the input power should vanish $\sigma_1 \to 0$ since $\sigma_1 \geq 0$.
It is clear that this should be the case since we have already noted that in general 
$\sigma_2 \to 0$ as $\eta \to 1$, therefore given the definition of $\eta$, $\sigma_1 \to 0$ as $\eta \to 1$.

\begin{figure}[ht]
  \centering  \includegraphics[width=\columnwidth]{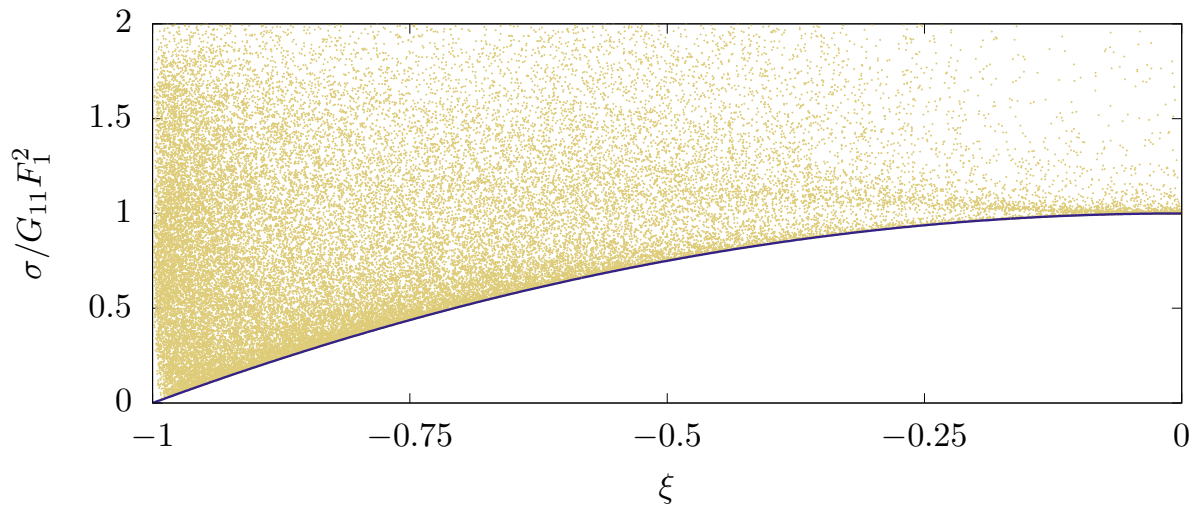}
  \caption{Total entropy production as function of the degree of coupling for the molecular motor model 
introduced in section \ref{sec:molec-motor-model}. 
The violet line is the bound of Eq.~(\ref{eq:boundsEP11}). 
The kinetic parameters of the model are randomly chosen
by multiplying the values used in Fig.~\ref{fig:coefficients}a by $e^x$ with $x$ drawn uniformly within $[-2, 2]$, 
whereas the values of the affinity are uniformly drawn within $[-3, 1]$ for $f$ and $[3,7]$ for $\Delta \mu$.}
  \label{fig:minimunEP}
\end{figure}

The improved bound on the total entropy production of \eqref{eq:boundsEP11}
is tested in Fig. \ref{fig:minimunEP} for a stochastic model of a molecular motor which will be presented in 
details in section \ref{sec:molec-motor-model}. The test consists in varying systematically kinetic parameters 
of the model and evaluating in each case the entropy production and the degree of coupling. The same figure for the bound \eqref{eq:boundsEP22} presents similar features but is not presented.
A related test also performed in the same way with this model checked that the maximum efficiency 
only depends on the degree of coupling \cite{Vroylandt2018}.

\section{Hierarchy of matrix inequalities}

\subsection{Conductance, activity and covariance matrices for Markov jump processes}
\label{sec:discr-syst-fram}

We use a Markov jump process to model a mesoscopic machine with a finite number of states. 
The probability per unit time to jump from state $y$ to state $x$ is given by the rate matrix $\bm{k}$ of 
components $k_{\exy} \geq 0$. We call the couple of states $\exy$ an oriented edge when $k_{\exy} > 0$. 
We assume that if the jump from $y$ to $x$ is possible then the reverse jump also exists, i.e. $k_{\exy} > 0$ 
implies that $k_{\eyx} > 0$.  The stationary probability of $x$, denoted $\pi_{x}$, verifies by definition 
$\sum_{y} k_\exy \pi_{y} = 0$. The mean probability current along edge $\exy$ in the stationary state is
\begin{equation}
  \label{eq:EdgeCurrent}
	\Jmoyi{\exy} \equiv k_{\exy} \pi_{y} - k_{\eyx} \pi_{x},
\end{equation}
 and the corresponding edge affinity writes 
\begin{equation}
  \label{eq:Edgedef}
  F_{\exy} \equiv \ln \f{k_{\exy} \pi_{y}}{ k_{\eyx} \pi_{x}}.
\end{equation}
We also introduce the physical matrix $\bm{\bar \phi}$ that connects the edge current to the physical current by
\begin{equation}
  \label{eq:physicalMat}
  J_\X = \sum_\exy \bar \phi_{\X,\exy} \Jmoyi{\exy}, \quad \X=\I,\,\II.
\end{equation}

From the mean probability currents and edge affinities, we define an edge conductance 
$\bar G_{\exy} \equiv \Jmoyi{\exy}/F_{\exy} $ which is a diagonal matrix in the space of edges. 
In \cite{Vroylandt2018}, we derived a unique expression of the non-equilibrium conductance matrix 
by combining edge resistances (inverse of edge conductances) in series, and cycle conductance in parallel, leading to 
\begin{equation}
    \label{eq:NonEqConducmatrixcomplet}
	\bm{G} \equiv\bm{\bar \phi}\cdot \bm{\mathcal{C}} \cdot\left(  \bm{\mathcal{C}}^{\T}  \cdot \bm{\bar G}^{-1}  \cdot \bm{\mathcal{C}} \right)^{-1}\cdot \bm{\mathcal{C}}^\mathrm{T} \cdot  \bm{\bar \phi}^\mathrm{T},
\end{equation}
where $\bm{\mathcal{C}}$ is the cycle matrix whose columns represent fundamental cycles on the graph of the machine 
and lines correspond to edges on the graph \cite{Polettini2016_vol94a, Polettini2016_vol94}. Each component of the matrix 
$\bm{\mathcal{C}}$ is $1$ or $-1$ if the edge belongs to the cycle (with sign + if the cycle and edge have the same
 orientation), and $0$ otherwise. 

In the study of non-equilibrium processes, the edge activity matrix $\bm{\bar A}$ of diagonal components
\begin{equation}
  \label{eq:EdgeActivity}
  \bar A_\exy  \equiv k_{\exy} \pi_{y} + k_{\eyx} \pi_{x},
\end{equation}
is of fundamental importance \cite{Maes2008a, Baiesi2009a, Basu2015, Wynants2010}. In this equation, 
$\bar{A}_{\exy}$ represents the mean number of jumps (irrespective of the direction of the jumps) 
per unit time between states $x$ and $y$ in the stationary state. 
The edge activity has a crucial influence on the edge resistance because if the machine almost never 
performs a transition along an edge (which means it has a low activity), this edge resistance should be high. 
This argument explains why the dynamical activity should matter not only for the thermodynamic uncertainty relations \cite{DiTerlizzi2018}, but more generally 
for key properties of the machine such its output power or its efficiency.
In exact parallel with the conductance matrix, we introduce the matrix of dynamical activity $\bm{A}$ as
\begin{equation}
    \label{eq:ActivityMatrix}
	\bm{A} \equiv \bm{\bar \phi}\cdot \bm{\mathcal{C}} \cdot\left(  \bm{\mathcal{C}}^{\T}  \cdot \bm{\bar A}^{-1}  \cdot \bm{\mathcal{C}} \right)^{-1}\cdot \bm{\mathcal{C}}^\mathrm{T} \cdot  \bm{\bar \phi}^\mathrm{T},
\end{equation}
where the edge activity appears instead of the edge conductance with respect to Eq.~(\ref{eq:NonEqConducmatrixcomplet}).

Finally, we define the covariance matrix $\bm{C}$ of physical currents
\begin{equation}
  \label{eq:cov_def}
  C_{\X,\Y} \equiv \lim_{t \to \infty} t\left[  \left \langle \Jstochi{\X} \Jstochi{\Y}  \right \rangle - \left \langle \Jstochi{\X} \right \rangle \left \langle \Jstochi{\Y}  \right \rangle \right],
\end{equation}
where $j_{\X}$ is the stochastic current for the driving process ($\X = 1$) or the output current ($\X = 2$). We denote by $\langle ... \rangle$ the mean value in the stationary state, i.e. $\langle j_{\X} \rangle = J_{\X} $. The covariance matrix characterizes the small fluctuations of currents around their average. 

Close to equilibrium case, the fluctuations-dissipation theorem connects the fluctuations characterized 
by the matrix $\bm{C}$ and the Onsager response matrix that is linked to dissipation. 
Far from equilibrium, the thermodynamic uncertainty relation replaces the fluctuations-dissipation theorem. 
In our framework, this shows up as a hierarchy of inequality for the matrices $\bm{G}$, $\bm{A}$ and $\bm{C}$, 
emphasizing the key role played by dynamical activity in non-equilibrium systems.

\subsection{From matrix inequalities to power-efficiency trade-offs}
\label{sec:few-inter-matr}

In order to compare the various matrices introduced above, it is useful to introduce among them
the Loewner partial order \cite{Book_Horn1985}. Given two symmetric 
$n\times n$ matrices $\bm{V}$ and $\bm{W}$, we write $\bm{V} \geq \bm{W}$ when $\bm{V}-\bm{W}$ is a 
positive semi-definite matrix, which also means that 
\begin{equation}
  \label{eq:loewnerpartialorder}
  \bm{V} \geq \bm{W} \Leftrightarrow \left( \forall \bm{x} \in  \mathbb{R}^n, \quad  \bm{x}^{\rm{T}} \cdot \bm{V} \cdot 
\bm{x} \geq \bm{x}^{\rm{T}} \cdot \bm{W} \cdot \bm{x} \right).
\end{equation}
With this definition, we derive in appendix \ref{App:IneqDerivation} the following matrix inequalities
using a large deviation framework :
\begin{equation}
  \label{eq:ineq_mat}
  \bm{G} \leq \f{\bm{A}}{2}  \leq \f{ \bm{C}}{2}.
\end{equation}
We view Eq.~(\ref{eq:ineq_mat}) as a fluctuation-activity-dissipation inequality.
At equilibrium, the non-equilibrium conductance matrix becomes the Onsager matrix 
$\bm{L}$, and the two inequalities above saturate because $\bm{L}=\bm{A}/2=\bm{C}/2$.

From Eqs.~(\ref{eq:loewnerpartialorder}) and (\ref{eq:ineq_mat}) and chosing $\bm{x} = (1,0)^{\rm T}$, we find
\be
G_{11} \leq \frac{1}{2} A_{11} \leq \frac{1}{2} C_{11}.
\ee
After multiplying these inequalities by $F_1^2$, we obtain $ G_{11} F_1^2 \leq A_{11} F_1^2/2 \leq C_{11} F_1^2/2$. 
Then, three different bounds on the output entropy production rate follows from Eq.~(\ref{eq:PowerEffS1}), 
in terms of the first coefficients of the non-equilibrium conductance, of the activity or of the current covariance matrices. 
\begin{equation}
  \label{eq:PowerEffTradeOff}
-\sigma_{\II} \leq  G_{11} F_\I^2 \eta (1-\eta) \leq  \frac{A_{11}}{2} F_\I^2 \eta (1-\eta) 
\leq  \frac{C_{11}}{2} F_\I^2 \eta (1-\eta),
\end{equation}
Note that Eq.~(\ref{eq:PowerEffTradeOff}) contains the trade-off derived by Pietzonka 
et al. \cite{Pietzonka2016b}.

In contrast to that, the trade-offs obtained by Sasa-Dechant\cite{Dechant2018}, see also Shiraishi 
\textit{et al.}\cite{Shiraishi2016} take the following form for Markovian dynamics on a graph: 
\begin{equation}
  \label{eq:orderingDS}
-\sigma_{\II} \leq   \frac{A_{11}}{2} F_\I^2 \eta (1-\eta) 
\leq  \f{1}{2} \Aphi F_\I^2 \eta (1-\eta),
\end{equation}
where $\Aphi= \sum_\exy \bar\phi_{\I,\exy}^2 A_\exy$ is an average dynamical activity   
with respect to the same function $\bar\phi_{\I,\exy}$ introduced in Eq.~\eqref{eq:physicalMat} to relate
physical and edge currents. Despite a common origin among all these trade-offs (see   
appendix~\ref{sec:avoid-direct-minim} for details), we note that there is no general 
ordering between $\Aphi$ in Eq.~(\ref{eq:orderingDS}) and the term proportional to $C_{11}$ 
in Eq.~(\ref{eq:PowerEffTradeOff}). 

We conclude this section by emphasizing that we focused on the bounds following from Eq.~\eqref{eq:PowerEffS1} combined with the matrix inequality of Eq.~\eqref{eq:ineq_mat} or with the bound for $A_{11}$ following from Cauchy-Schwartz inequality, but it is straightforward to obtain similar upper bounds for the other inequalities in section \ref{general-discussion}.

\section{Illustrative examples}
In this section, we illustrate the above power-efficiency bounds using two simple models of thermodynamic autonomous machines studied in Ref.~\cite{Vroylandt2018} : a unicyclic thermal engine and an isothermal molecular motor that has several cycles. We first describe these two models and then discuss our main results.

\subsection{Unicyclic thermal engine}
\label{sec:unicycl-therm-engine}

We start with the unicyclic heat-to-heat converter with three states $a,b$ and $c$ of energy $E_a,E_b,E_c$. Each transition is promoted by a different heat reservoir at inverse temperature $\beta_1,\beta_2,\beta_3$. We take the Boltzmann constant $k_{B} =1 $, and set the energy scale by taking $\beta_{3} = 1$. The transition rates are
\begin{equation}
  \label{eq:ratesUE}
  \begin{array}{ll}
    k_{(b,a)}= \Gamma e^{-\f{\beta_1}{2}(E_b-E_a)},&k_{(a,b)}= \Gamma e^{-\f{\beta_1}{2}(E_a-E_b)},\\
    k_{(c,b)}= \Gamma e^{-\f{\beta_2}{2}(E_c-E_b)},&k_{(b,c)}= \Gamma e^{-\f{\beta_2}{2}(E_b-E_c)},\\
    k_{(a,c)}= \Gamma e^{-\f{\beta_3}{2}(E_a-E_c)},&k_{(c,a)}= \Gamma e^{-\f{\beta_3}{2}(E_c-E_a)},\\
  \end{array}
\end{equation}
where $\Gamma$ is the coupling constant to the heat reservoirs which defines the unit of time and which we take to be 
$\Gamma = 1$. Since the converter is coupled to three heat reservoirs, the total entropy production rate writes 
$\sigma = -\beta_1 \Jmoyi{1} - \beta_2 \Jmoyi{2} - \beta_3 \Jmoyi{3}$, where $\Jmoyi{i}$ denotes the heat flux 
from the heat reservoir $i$ to the system. Using energy conservation $\Jmoyi{1}+\Jmoyi{2}+\Jmoyi{3}=0$, we simplify 
the total entropy production rate as $\sigma =  (\beta_3-\beta_1)\Jmoyi{1} + (\beta_3-\beta_2)\Jmoyi{2}$. In agreement 
with section \ref{general-discussion}, we consider as driving process the heat flow $\Jmoyi{1}$ and output process the
 heat flow $\Jmoyi{2}$. Without loss of generality, we assume the following inequalities for the reservoir's temperatures
 $\beta_3 > \beta_1$ and $\beta_3>\beta_2$ and for the energy levels $E_b> E_c > E_a$. 
Under these conditions, the driving and output currents are such that $\Jmoyi{1}>0$ and $\Jmoyi{2}<0$: 
the system operates as a machine that transfers heat 
from a cold to a hot reservoir using the thermodynamic force generated by the transfer of heat from a hot to a cold reservoir.
The partial entropy production rates and physical affinities are then
\begin{equation}
  \label{eq:PhysicalUE}
  \begin{array}{ll}
    \sigma_1 = (\beta_3-\beta_1)\Jmoyi{1},& F_1=(\beta_3-\beta_1)\\
     \sigma_2 = (\beta_3-\beta_1)\Jmoyi{2},& F_2=(\beta_3-\beta_2).
  \end{array}
\end{equation}

We emphasize that this model is unicyclic and hence satisfies the tight coupling condition. 
Therefore, the currents $\Jmoyi{1}$ and $\Jmoyi{2}$ are proportional to each other and at stalling, \textit{i.e.} 
when $\Jmoyi{2}=0$, the heat to heat converter works reversibly and does not produce entropy.

\subsection{Molecular motor model}
\label{sec:molec-motor-model}

Our second example is a discrete model of a molecular motor \cite{Lau2007_vol99,Lacoste2008_vol78}. 
The motor has only two internal states and evolves on a linear discrete lattice by consuming Adenosine TriPhosphate (ATP) molecules.
The position of the motor is given by two variables: the position $n$ on the lattice and $y$ is the number of ATP consumed. The even and odd sites are denoted by $a$ and $b$, respectively.  Note that the lattice of $a$ and $b$ sites extends indefinitely in both directions along the $n$ and $y$ axis; for the spatial direction $n$, the lattice step defines the unit length.
There are two physical forces acting on the motor, a chemical force controlled by the chemical potential difference of the hydrolysis reaction of ATP, $\Delta \mu$ and a mechanical force $f$ applied directly on the motor. The whole system is in contact with a heat bath, and we choose to express all quantities in units of $k_B T$. Equilibrium corresponds to the 
vanishing of the two currents, namely the mechanical current $\bar v$ which is the average velocity of the motor on the lattice,  
and the chemical current $r$, which is its average rate of ATP consumption. 
Since the system operates cyclically, the change of internal energy in a cycle is zero and the first law takes the form $q+r\Delta\mu+f \bar v=0$ where $q$ is the heat flow coming from the heat bath, $r \Delta \mu$ represents the chemical work and $f \bar v$ represents the mechanical work; all quantities are evaluated in a cycle. 
Under these conditions, the second law takes the form $\sigma=- q$, and 
the entropy production rate takes the following form:
\begin{equation}
  \label{eq:MMentProd}
  \sigma =  f \bar v + r \Delta \mu.
\end{equation}
In the normal operation of the motor, chemical energy
is converted into mechanical energy, which means that the driving process $(1)$ is the chemical one and the output  
process $(2)$ the mechanical one in agreement with the convention made in this paper. Thus, the two partial entropy 
production rates should be $\sigma_\I = r \Delta \mu$, with the chemical affinity $F_\I =  \Delta \mu$ and 
$\sigma_\II = f \bar v$, with mechanical affinity $F_\II = f$.

\begin{figure}[ht]
  \centering  \includegraphics[width=\columnwidth]{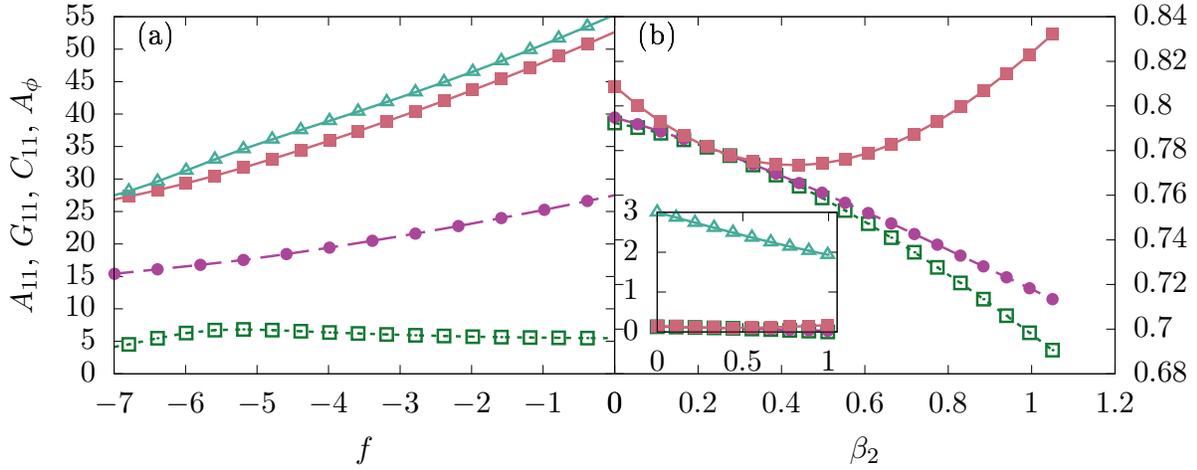}
  \caption{Trade-off coefficients versus the force acting on the molecular motor in (a) or 
versus the inverse temperature for the unicyclic thermal engine in (b). 
The Green empty squares are the coefficient $G_{11}$ of the conductance matrix, 
the violet circles are the coefficient $A_{11}$ of the activity matrix, the coefficient $C_{11}$ 
of the covariance matrix is shown with the red full squares and the input power activity $\Aphi$ is the 
blue empty triangles. Insert: A larger view of the main figure.
For figure (a), the parameters are $\Delta \mu =20.0$, $\alpha=0.57$, $\alpha'=1.3.10^{-6}$, $\omega=3.5$, 
$\omega'=108.15$ $\epsilon=10.81$, $\theta_a^+=0.25$, $\theta_a^-=1.83$, $\theta_b^+=0.08$, $\theta_b^-=-0.16$. 
For figure (b), they are  $\beta_1=0.5$, $\beta_3=1$, $\Gamma=1$, $E_a=1$,  $E_b=4$ and $E_c=2$.}
  \label{fig:coefficients}
\end{figure}
\begin{figure}[ht]
  \centering  \includegraphics[width=\columnwidth]{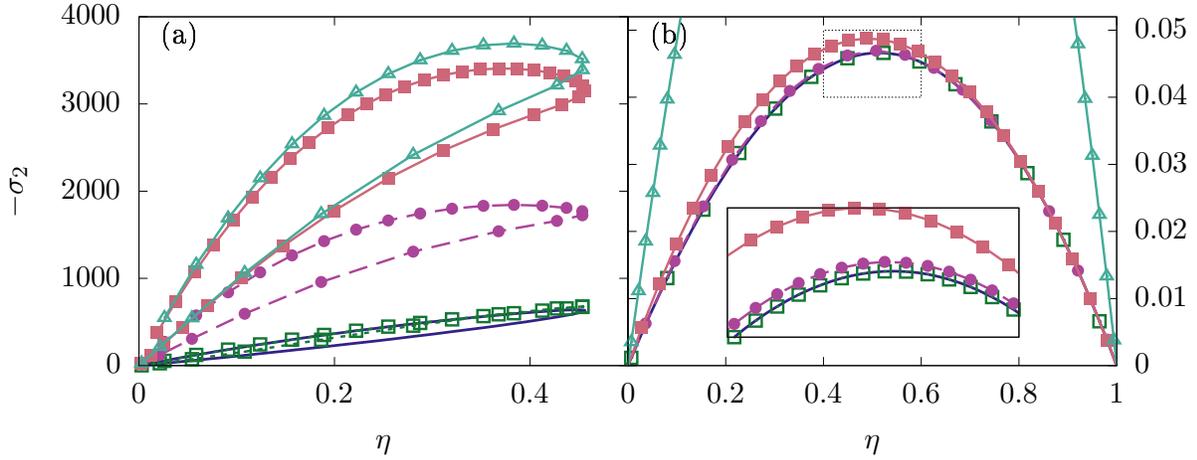}
  \caption{Output entropy production rate as a function of the thermodynamic efficiency for 
  the molecular motor (left) or the unicyclic thermal engine (right). The solid line represents the
 output entropy production and the symbols represent the different power-efficiency trade-offs 
derived from the coefficients represented in Fig.~\ref{fig:coefficients} (with the same color code and shape). 
Inset: Zoom in the region of the maximum power. Parameters are the same than in Fig.~\ref{fig:coefficients}.}
  \label{fig:power}
\end{figure}

\subsection{Discussion}
\label{sec:discussion}

In order to illustrate the inequalities (\ref{eq:ineq_mat}), we plot the $(1,1)$
coefficients of the three matrices $\bm{G}$, $\bm{A}/2$, $\bm{\text{Cov}}/2$ and the dynamical activity parameter 
$\Aphi$ in Fig.~\ref{fig:coefficients} for the unicyclic engine and the molecular motor as function of  
the output affinities for both machines. 
We confirm the order between the different coefficients predicted by 
Eq.~\eqref{eq:ineq_mat}-Eq.~\eqref{eq:orderingDS}.

In the chosen conditions, only 
the unicyclic engine can approach equilibrium, it does so around $\beta_2=0.3$.
At this point, all three coefficients converge towards the same value. 

On Fig.~\ref{fig:power}, we plot the four studied 
power-efficiency trade-offs. We confirm again the order among the four trade-offs. It can be also observed that in the case of the tight coupling machine, the inequality (\ref{eq:PowerEffS1}) is even an equality \cite{Vroylandt2018}.

It is interesting at this point to observe that the quality of the various bounds seems to be related to the level of information available about the system. Indeed, the tightest bound is the one obtained from the non-equilibrium conductance matrix, which is built using the knowledge of the microscopic dynamics of the system. The bound obtained from the dynamical activity is less tight, but it also requires less information since only time symmetric observables of the microscopic dynamics are used. The bounds deduced from the covariance matrix are the most loose bounds, and they indeed require the least information, since only information on macroscopic physical currents is needed instead of the more detailed stochastic dynamics of edge currents.

\section{Conclusion}
\label{sec:conclusion}
In this work, we have extended our previous framework on the conductance matrix for general thermodynamic machines operating in a non-equilibrium steady state arbitrarily far from equilibrium. By parametrizing this conductance matrix in terms of the degree of coupling, we obtain various bounds for the input and output power and for the total entropy production. It is easy to see that the bounds on the total or partial entropy production go beyond the second law of thermodynamics.

While these bounds can be proven generally, they involve a constant factor, a coefficient of the conductance matrix, which is in general unknown. To make progress, we choose a discrete Markov jump process for the microscopic dynamics, which allows to calculate explicitly important matrices for this problem, such as  the conductance matrix, the activity matrix and the covariance matrix. We show that these matrices are ordered according to Loewner partial order, and that these matrix inequalities contain an ordered set of power-efficiency trade-offs.

Our formulation includes a number of already known results such as the power-efficiency trade-off derived by Pietzonka and Seifert or the inequality previously obtained by Dechant and Sasa for Langevin systems. 
We obtain a hierarchy of power-efficiency trade-offs, with an order that depends primarily on the level of knowledge of the microscopic dynamics. The tightest bound is obtained when the maximum of information is available on the microscopic dynamics, while more loose bounds are obtained when only coarse-grained information is available.

The present work applies to stationary machines but not to periodically driven ones \cite{Koyuk2018,Barato2018b}.
We have also not considered systems with broken time-reversal symmetry \cite{Proesmans2019} for which 
extensions of this framework could be carried out. We hope to address some of these extensions in future work.

\appendix

\section{Derivation of the matrix inequalities}
\label{App:IneqDerivation}
Let us consider in a stochastic description of the machine, a long trajectory of 
duration $T$, and $x(t)$ the label of the state occupied at time $t$. The empirical density is defined as the fraction of time a given trajectory spends in state $y$ as
\be
p_y = \frac{1}{T} \int_{0}^{T} dt \delta_{x(t),y}.
\ee
In the long time limit, $p_y$ tends to $\pi_y$ which is the stationary probability distribution.
Furthermore, we denote the empirical edge current associated to the net number of transitions from $y$ to $x$ per unit time during a trajectory of duration $T$ by $\Jstochi{\exy}$, with
\be
\Jstochi{\exy}= \frac{1}{T} \int_{0}^{T} dt \left( \delta_{x(t^-),y} \delta_{x(t^+),x} - 
\delta_{x(t^+),y} \delta_{x(t^-),x} \right), 
\ee
where $x(t^\pm)$ denotes the configuration immediately before or after time $t$.
In the long time limit, $\Jstochi{\exy}$ tends to $J_\exy$, which is the steady state current. Beside these two currents, let's introduce the current $j^p_{\exy}$ that
represents the expected edge current given the empirical density $\bm{p}$ and the edge rates
\be
j^p_\exy = k_\exy p_\oe - k_\eyx p_\ome.
\ee
Finally, we denote by $g^p_\exy$ the edge rates given $\bm{p}$
that represents the pairwise geometric average on direction of each transition rate
\begin{eqnarray}
  \label{eq:defs}
  g^p_\exy &=& 2 \sqrt{k_\exy k_\eyx p_\oe p_\ome}. 
\end{eqnarray}

The probability distribution $P( \{p_x\}, \{ \Jstochi{\exy} \})$ of the empirical density and edge currents 
obeys at large time $T$ a large deviation principle yielding
\begin{equation}
  \label{eq:LDFprob}
  P( \{p_x\}, \{ \Jstochi{\exy} \}) \simeq e^{-T I( \{p_x\},
\{ \Jstochi{\exy} \} )  },
\end{equation}
where $I( \{p_x\},\{ \Jstochi{\exy} \} )$ is a 
large deviation function (LDF) \cite{Touchette2009_vol478}. 
This LDF provides the rate at which decays with time the probability that empirical densities and edge currents remain different from their steady state values. This level of description is called the level 2.5 in the literature. The LDF 
at that level for Markov jump processes has an explicit form \cite{Gingrich2017a}:
\begin{eqnarray}
  \label{eq:LDF2.2_various_form}
  I_{2.5}(\{p_x\},\{j_\exy\}) &=&   \sum_{\exy}  j_\exy \text{arcsinh}\left( \f{j_\exy}{{\geoact^p_\exy}} \right)- j_\exy \text{arcsinh}\left( \f{j^p_\exy}{{\geoact^p_\exy}} \right)\nonumber \\
                                   & &+\sqrt{ {j^p_\exy}^2+{\geoact^p_\exy}^2} -\sqrt{ j_\exy^2+{\geoact^p_\exy}^2}.
\end{eqnarray}
To make useful predictions based on this LDF one must coarse-grain edge currents into physical currents \cite{Verley2016_vol93}, using that the latter are linearly related to the formers. Hence, the LDF for physical currents is obtained from Eq.~(\ref{eq:LDF2.2_various_form}) by the following contraction:
\begin{equation}
  \label{eq:contractionI2.5}
  I(\vect{j}) = \min\limits_{\{p_x\},\{..\}} I_{2.5}(\{p_x\},\{j_\exy\}),
\end{equation}
where $\{..\}$ denotes here (and in the following) the minimum over edge currents $\{j_\exy\}$ that corresponds to the physical current $\vect{j}$ and respect the stationary condition
\begin{equation}
  \label{eq:statCOndLDF}
 \forall x ,\quad \sum_y( j_{\exy} -   j_{\eyx}) = 0.
\end{equation}
A first bound follows from Eq.~\eqref{eq:contractionI2.5}, once  
the empirical density $\{p_x\}$ is approximated 
by the stationary probability $\{\pi_x\}$, namely:
\begin{equation}
  \label{eq:contractionI2.5pi}
  I(\vect{j}) \leq \min\limits_{\{..\}} I_{2.5}(\{\pi_x\},\{j_\exy\}).
\end{equation}
By performing a Taylor expansion of $I_{2.5}(\{\pi_x\},\{j_\exy\})$ around $j_\exy \simeq J_\exy$ at second order, 
one obtains an approximated function which we call
$I_\text{loc}(\{\pi_x\},\{j_\exy\})$, with
\begin{equation}
  \label{eq:approxLDF2.5}
  I_\text{loc}(\{\pi_x\},\{j_\exy\}) = \sum_{\exy} \f{(j_\exy - J_\exy )^2}{2\sqrt{ {J_\exy}^2+{\geoact^\pi_\exy}^2}} =  \sum_{\exy} \f{(j_\exy - J_\exy )^2}{2 \act_\exy}
\end{equation}
Therefore, combining Eq.~(\ref{eq:contractionI2.5pi}) and (\ref{eq:approxLDF2.5}) leads to the local bound on current LDF 
\begin{equation}
  \label{eq:bornes_loc}
  I(\vect{j}) \leq \min\limits_{\{..\}} I_{2.5}(\{\pi_x\},\{j_\exy\}) \simeq  \min\limits_{\{..\}} I_\text{loc}(\{\pi_x\},\{j_\exy\}).
\end{equation}
We emphasize that the Eq.~(\ref{eq:bornes_loc}) is a local bound in the sense that it is valid only up to the second order of the Taylor expansion. As shown in Ref.~\cite{Gingrich2017a}, a closely related bound 
denoted $I_{quad}$ lead this time to a global bound, namely
$I_{2.5}(\{\pi_x\},\{j_\exy\}) \leq I_\text{quad}(\{\pi_x\},\{j_\exy\})$, with
\begin{equation}
  \label{eq:bornes_quad_LDF}
  I_\text{quad}(\{\pi_x\},\{j_\exy\}) = \f{1}{4} \sum_{\exy} (j{\exy} -  J_{\exy} )^2 
\frac{\sigma^\pi_\exy}{J_{\exy}^2}.
\end{equation}
In this equation, $\sigma^\pi_\exy$ is the steady state entropy production rate associated to the transitions from $y$ to $x$ defined by
\be
\sigma^\pi_\exy=( k_\exy \pi_\oe - k_\eyx \pi_\ome ) \ln \frac{k_\exy \pi_\oe}{k_\eyx \pi_\ome}.
\ee
Now, using the relation $\sigma^\pi_\exy=J_\exy F_\exy$ and the 
definition $\bar{R}_\exy=F_\exy/J_\exy$, one can write $I_\text{quad}$ as
\be
I_\text{quad}(\{\pi_x\},\{j_\exy\}) =  
\f{1}{4} \sum_{\exy} (j{\exy} -  J_{\exy} )^2 \bar{R}_\exy.
\ee
Further, using the general inequality $(a-b) \ln (a/b) \geq 2 (a-b)^2 /(a+b)$, one deduces first that $\sigma^\pi_\exy  \geq 2 J_{\exy}^2/ \act_\exy  $ and then using Eq.~\eqref{eq:bornes_quad_LDF} that
\be
\label{somebound}
I_\text{loc}(\{\pi_x\},\{j_\exy\}) \leq  I_\text{quad}(\{\pi_x\},\{j_\exy\}).
\ee
Using Eqs.~\eqref{eq:contractionI2.5},\eqref{eq:bornes_loc} and \eqref{somebound}, we obtain in the end:
\begin{equation}
  \label{eq:sucess_bornes_loc}
   I(\vect{j}) \leq  \min\limits_{\{..\}} 
I_\text{loc}(\{\pi_x\},\{j_\exy\}) \leq   \min\limits_{\{..\}}  I_\text{quad}(\{\pi_x\},\{j_\exy\}).
\end{equation}
Since we are now minimizing quadratic functions, we can find the minimizer exactly as in Ref.~\cite{Vroylandt2018}:
\begin{eqnarray}
  \label{eq:minimizationA}
  I_\text{loc}(\vect{j})&= \min\limits_{\{..\}} I_\text{loc}(\{\pi_x\},\{j_\exy\})=&  \frac{1}{2} 
\left( \vect{j} - \vect{J} \right)^{\T} \cdot 
\bm{A}^{-1} \cdot \left( \vect{j} - \vect{J} \right) \\\label{eq:minimizationG}
 I_\text{quad}(\vect{j})&=  \min\limits_{\{..\}}  I_\text{quad}(\{\pi_x\},\{j_\exy\}) =&  
\frac{1}{4} \left( \vect{j} - \vect{J} \right)^{\T} \cdot 
\bm{G}^{-1} \cdot \left( \vect{j} - \vect{J} \right)
\end{eqnarray}
with the expression of the matrix $\bm{G}$ and $\bm{A}$ being given by the Eqs.~(\ref{eq:NonEqConducmatrixcomplet}) and~(\ref{eq:ActivityMatrix}). Since
\begin{equation}
\fl  I(\vect{J}) = I_\text{loc}(\vect{J}) = I_\text{quad}(\vect{J}) =0 
\quad\text{and}\quad \frac{dI}{dj}(\vect{J}) = \frac{dI_\text{loc}}{dj}(\vect{J}) = \frac{dI_\text{quad}}{dj}(\vect{J}) =0,
\end{equation}
the inequality~\eqref{eq:sucess_bornes_loc} propagates to second order derivatives:
\begin{equation}
  \label{eq:ineq_matrix_inv}
 \bm{C}^{-1} \leq  \bm{A}^{-1} \leq \f{1}{2} \bm{G}^{-1}.
\end{equation}
Using properties of semi-definite positive matrices \cite{Book_Horn1985} ends the proof of Eq.~\eqref{eq:ineq_mat}
\begin{equation}
  \label{eq:ineq_mat-2}
  \bm{G} \leq \f{\bm{A}}{2}  \leq \f{\bm{C}}{2}.
\end{equation}

\section{Bound from an activity ansatz}
\label{sec:avoid-direct-minim}

The computation of conductance matrix and activity matrix in Eq.~(\ref{eq:minimizationA}--\ref{eq:minimizationG}) 
requires the minimization of the bounds. Instead, we can rely on the use of an ansatz if we focus on only one current. Let's consider the stochastic current $j_1$ defined as a linear combination of edge currents
\begin{equation}
  \label{eq:generalizedCurrent}
 j_\I = \sum_{\exy} \bar \phi_{\I,\exy} j_{\exy}.
\end{equation}
To avoid the minimization in Eq.~(\ref{eq:sucess_bornes_loc}), we use an ansatz on edge current $\tilde j_\exy(j_1)$ that verifies
\begin{equation}
  \label{eq:ansatzCond}
   \sum_{\exy} \bar\phi_{\I,\exy} \tilde j_\exy(j_1) = j_1,
\end{equation}
and the stationary condition
\begin{equation}
  \label{eq:ansatzStatCond}
  \forall x ,\quad \sum_y( \tilde j_{\exy}(j_1) -  \tilde j_{\eyx}(j_1)) = 0.
\end{equation}
Following Ref.\cite{Gingrich2016_vol116}, the ansatz 
\begin{equation}
  \label{eq:ansatz1}
  \tilde j_{\exy}(j_1) = J_{\exy} \f{j_1}{J_1}.
\end{equation}
works and can be used into Eq.~\eqref{eq:sucess_bornes_loc} yielding
\begin{equation}
  \label{eq:ineqWithAnsatz}
  I(j_1) \leq I_\text{loc}(\{\pi_x\},\{ \tilde j_{\exy}(j_1)\}) \leq  I_\text{quad}(\{\pi_x\},\{ \tilde j_{\exy}(j_1)\}).
\end{equation}
The second derivative of this equation with respect to $j_1$ leads to
\begin{equation}
  \label{eq:ineqWithAnsatzUncert}
\f{1}{ \mbox{Var}(j_1)} \leq \f{\left( \sum_\exy \f{J_\exy^2}{\act_\exy}\right)}{J_1^2}\leq \f{\sigma}{2J_1^2} .
\end{equation}
Due to Cauchy-Schwarz inequality, we have
\begin{equation}
  \label{eq:CSineqDS}
 J_1^2 = \left( \sum_\exy \bar \phi_{\I,\exy} J_\exy\right)^2\leq \left( \sum_\exy \bar\phi_{\I,\exy}^2 \act_\exy\right) \left( \sum_\exy \f{J_\exy^2}{\act_\exy}\right),
\end{equation}
where actually $\act_{\exy}$ could be arbitrary.
Combining Eq.~(\ref{eq:ineqWithAnsatzUncert}) and~(\ref{eq:CSineqDS}) gives then
\begin{equation}
  \label{eq:CSonact}
  \f{J_1^2}{\sigma} \leq \f{J_1^2}{2\left( \sum_\exy \f{J_\exy^2}{\act_\exy}\right)}\leq \f{1}{2}\left( \sum_\exy \bar\phi_{\I,\exy}^2 \act_\exy\right) = \f{1}{2} \Aphi.
\end{equation}
This equation is similar to the bound derived for Langevin systems in Ref.\cite{Dechant2018} (see Eq.~(14) of that reference). It expresses a bound on the square of any current (here $J_1^2$) in terms of the total entropy production times a coefficient which depends on the activity. In diffusive systems, this activity may be expressed in terms of the diffusion coefficient of the system. We note that while the linear decomposition of Eq.~\eqref{eq:generalizedCurrent} is general, we need to choose the specific function $\bar\phi_{\I,\exy}$ introduced in Eq.~\ref{eq:physicalMat} in order to apply the Cauchy-Schwartz inequality specifically to physical currents.

Notice that the term in the rhs of Eq.~(\ref{eq:CSonact}) could also be obtained by using as ansatz
\begin{equation}
  \label{eq:ansatzBis}
  \tilde j_\exy (j_1) = J_\exy + (j_1-J_1) \f{\bar \phi_{\I,\exy} \act_\exy}{\sum_\exy \bar\phi_{\I,\exy}^2 \act_\exy},
\end{equation}
that respect the condition~(\ref{eq:ansatzCond}) but not the stationary 
condition~(\ref{eq:ansatzStatCond}). It happens that the ansatz~(\ref{eq:ansatzBis}) 
is the actual miminizer of $ I_\text{loc}(\{\pi_x\},\{j_\exy\})$ under the constraint~(\ref{eq:ansatzCond}) but without considering the stationary condition. Therefore, pluging the ansatz of Eq.~(\ref{eq:ansatzBis}) inside $  I_\text{loc}(\{\pi_x\},\{ j_\exy \})$ leads to
  \begin{equation}
    \label{eq:orderingI_loc}
  I_\text{loc}(\{\pi_x\},\{ \tilde j_\exy (j_1)\}) =  \f{ (j_1-J_1)^2}{2\Aphi} \leq \min\limits_{\{..\}}  I_\text{loc}(\{\pi_x\},\{j_\exy\}),
  \end{equation}
where as before the minimum of right hand side is carried over $\{j_\exy\}$ that corresponds to physical current $j_1$ and respect the stationary condition (\ref{eq:ansatzStatCond}). Hence, using Eq.~(\ref{eq:minimizationA}) and by deriving twice with respect to $j_{1}$, we obtain the inequality $A_{11} \leq \Aphi $ used in Eq.~(\ref{eq:orderingDS}).

\section{Illustrative example: conductance and activity matrices}
\label{sec:expression-complete}

\subsection{Unicyclic heat-to-heat converter}
\label{sec:unicyclic-heat-heat}
Given the rates of the unicyclic heat-to-heat converter, we are able to determine the stationary probabilities $\pi_a,\, \pi_b$ and $\pi_c$ using for instance the spanning tree formula. We next compute the stationary cycle current $\Jmoyi{c_1}$ and the mean activity on each edge $(i,j)$
\begin{equation}
  \label{eq:meanact_unicyclic}
  \bar A_{(i,j)} = k_{(j,i)}\pi_i+ k_{(i,j)}\pi_j
\end{equation}
That give us the conductance 
\begin{equation}
  \label{eq:Gijunicyclic}
\fl  \bm{G} =\f{\Jmoyi{c_1}}{F_{c_1}}
  \left( \begin{array}{cc}
    (E_b-E_a)^2 & (E_c-E_b)(E_b-E_a) \\ (E_c-E_b)(E_b-E_a) & (E_c-E_a)^2\\
  \end{array} \right),
\end{equation}
and activity matrix
\begin{equation}
  \label{eq:Aijunicyclic}
\fl  \bm{A} =\left(\f{1}{\bar A_{(a,b)}}+\f{1}{\bar A_{(b,c)}}+\f{1}{\bar A_{(c,a)}} \right)^{-1}
  \left( \begin{array}{cc}
    (E_b-E_a)^2 & (E_c-E_b)(E_b-E_a) \\ (E_c-E_b)(E_b-E_a) & (E_c-E_a)^2\\
  \end{array} \right).
\end{equation}
These expression are used to draw Fig.~\ref{fig:coefficients}b and \ref{fig:power}b.

\subsection{MolecularMotor}
\label{sec:molecularmotor}
The graph of this model includes four bidirectional edges connecting two states.  
For two of these edges, the transitions are passive and do not consume or produce ATP, but the two others are active. 
The eight transition rates associated to these four bidirectional edges are 
\begin{equation}
  \begin{array}{ll}
\overrightarrow{\omega_b}^{-1}= \alpha' e^{\theta^{+}_b f}, &\overrightarrow{\omega_b}^{0} =  \omega'\,e^{\theta^{+}_b f},\\
 \overleftarrow{\omega_a}^{1}  =  \alpha' e^{ -\epsilon + \Delta \mu - \theta^{-}_a f}, &  \overleftarrow{\omega_a}^{0} = \omega'\,e^{-\epsilon- \theta^{-}_a f},\\
  \overleftarrow{\omega_b}^{-1}= \alpha\, e^{-\theta^{-}_b f}, & \overleftarrow{\omega_b}^{0} =  \omega\,e^{-\theta^{-}_b f},  \\
 \overrightarrow{\omega_a}^{1}  =  \alpha\, e^{ -\epsilon + \Delta \mu + \theta^{+}_a f }, & \overrightarrow{\omega_a}^{0} = \omega\,e^{-\epsilon + \theta^{+}_a f} \, ,
\end{array}
\label{eq:ratesMolecularMotor}
\end{equation}
where we have kept the original notation of Refs.~\cite{Lau2007_vol99,Lacoste2008_vol78}. In the above equations,  $\theta_i^\pm$ represent load distribution factors that are arbitrary except that $\theta^{+}_a+\theta^{-}_b + \theta^{-}_a+\theta^{+}_b\ =2~$\cite{Lacoste2008_vol78}. Let's orientate all edges from state $a$ to $b$. Then, the four edge currents and affinities are
\ba
\label{eq:MMedeLabel-1}
J_{(1)} &= \pi_a \overleftarrow{\omega_a}^{1} - \pi_b \overrightarrow{\omega_b}^{-1}, 
& \qquad F_{(1)}= \ln \frac{\overleftarrow{\omega_a}^{1}\pi_{a}}{\overrightarrow{\omega_b}^{-1}\pi_{b}}, \\
J_{(2)} &= \pi_a \overleftarrow{\omega_a}^{0} - \pi_b \overrightarrow{\omega_b}^{0}, 
& \qquad F_{(2)}= \ln \frac{\overleftarrow{\omega_a}^{0}\pi_{a}}{\overrightarrow{\omega_b}^{0}\pi_{b}}, \\
J_{(3)} &= \pi_a \overrightarrow{\omega_a}^{0} - \pi_b \overleftarrow{\omega_b}^{0}, 
& \qquad F_{(3)}= \ln \frac{\overrightarrow{\omega_a}^{0}\pi_{a}}{\overleftarrow{\omega_b}^{0}\pi_{b}}, \\
J_{(4)} &= \pi_a \overrightarrow{\omega_a}^{1} - \pi_b \overleftarrow{\omega_b}^{-1},
& \qquad F_{(4)}= \ln \frac{\overrightarrow{\omega_a}^{1}\pi_{a}}{\overleftarrow{\omega_b}^{-1}\pi_{b}},
\label{eq:MMedgeLabel-4}
\ea
in terms of the stationary probabilities of states $a$ or $b$, denoted $\pi_a$ and $\pi_b$ respectively. For the explicit expressions of the probability currents in terms of the transition rates, we refer to Ref.~\cite{Lau2007_vol99,Lacoste2008_vol78}. If one introduce the edge resistance matrix $\bar R_{(i)} = F_{(i)}/J_{(i)}$ with $i=1,2,3$ and $4$, the conductance matrix for this model writes
\begin{equation}
  \label{eq:Gij-molecular}
\fl \bm{G} =\f{1}{Z_G}  \left( \begin{array}{cc}
  ( \bar R_{(1)}+ \bar R_{(4)} )(\bar R_{(3)}+ \bar R_{(2)}) & 2(\bar R_{(4)}   \bar R_{(2)}-\bar R_{(1)}  \bar R_{(3)}) \\
 2(\bar R_{(4)}   \bar R_{(2)}-\bar R_{(1)}  \bar R_{(3)})&  4( \bar R_{(1)}+ \bar R_{(2)} )(\bar R_{(3)}+ \bar R_{(4)}) 
  \end{array} \right), 
\end{equation}
with
\begin{equation}
  \label{eq:nomrFactormatrixGMM}
 Z_G  = \bar R_{(1)} \bar R_{(4)}  \bar R_{(3)} +  \bar R_{(1)} \bar R_{(4)}  \bar R_{(2)} +  \bar R_{(1)} \bar R_{(3)} \bar R_{(2)} + \bar R_{(4)}  \bar R_{(3)} \bar R_{(2)}.
\end{equation}
The activity matrix is derived in a similar way and we obtain
\begin{equation}
  \label{eq:Aij-molecular}
\fl \bm{A} =\f{1}{Z_A}  \left( \begin{array}{cc}
  ( \bar A^{-1}_{(1)}+ \bar A^{-1}_{(4)} )(\bar A^{-1}_{(3)}+ \bar A^{-1}_{(2)}) & 2(\bar A^{-1}_{(4)}   \bar A^{-1}_{(2)}-\bar A^{-1}_{(1)}  \bar A^{-1}_{(3)}) \\
 2(\bar A^{-1}_{(4)}   \bar A^{-1}_{(2)}-\bar A^{-1}_{(1)}  \bar A^{-1}_{(3)})&  4( \bar A^{-1}_{(1)}+ \bar A^{-1}_{(2)} )(\bar A^{-1}_{(3)}+ \bar A^{-1}_{(4)}) 
  \end{array} \right), 
\end{equation}
with
\begin{equation}
  \label{eq:nomrFactormatrixAMM}
 Z_A  = \bar A^{-1}_{(1)} \bar A^{-1}_{(4)}  \bar A^{-1}_{(3)} +  \bar A^{-1}_{(1)} \bar A^{-1}_{(4)}  \bar A^{-1}_{(2)} +  \bar A^{-1}_{(1)} \bar A^{-1}_{(3)} \bar A^{-1}_{(2)} + \bar A^{-1}_{(4)}  \bar A^{-1}_{(3)} \bar A^{-1}_{(2)}.
\end{equation}
and
\ba
\label{eq:MMedeLabel-1}
A_{(1)} &=& \pi_a \overleftarrow{\omega_a}^{1} + \pi_b \overrightarrow{\omega_b}^{-1}, \quad
A_{(2)} = \pi_a \overleftarrow{\omega_a}^{0} + \pi_b \overrightarrow{\omega_b}^{0},  \\
A_{(3)} &=& \pi_a \overrightarrow{\omega_a}^{0} + \pi_b \overleftarrow{\omega_b}^{0},   \quad \; \;
A_{(4)} = \pi_a \overrightarrow{\omega_a}^{1} + \pi_b \overleftarrow{\omega_b}^{-1}. 
\label{eq:MMedgeEdgeActivity}
\ea

\section*{References}

\bibliographystyle{unsrt}
\bibliography{Tradeoff}

\end{document}